\documentclass[prl,twocolumn,showpacs,preprintnumbers,amsmath,amssymb,euscript,amsfonts]{revtex4}
\usepackage{graphicx}
\usepackage{amssymb}
\usepackage{dcolumn}
\usepackage{bm} 
\usepackage[mathscr]{eucal}
\usepackage{color}
\usepackage{subfigure}
\usepackage{amsmath}

\newcommand{\ie}{{\it i.e.}}

\def\Hlocal#1{ H_{\sigma}^{_{(#1)}} }
\def\pdagger#1{ p_{#1,\sigma}^{_{(i)} \dagger} }
\def\tpdagger#1{ \tilde{p}_{#1,\sigma}^{_{(i)} \dagger} }
\def\opd#1{ d_{#1,\sigma}^{_{ (i)}} }

\def\ip#1{p^{_{(i)}}_{#1} }
\def\tipp#1{\tilde{p}^{_{(i+1)}}_{#1}}
\def\id#1#2{ d_{#1,#2}^{_{ (i)}}}

\def\iB#1#2{B_{#1#2}^{_{(i)}}}
\def\iBp#1#2{B_{#1#2}^{_{(i+1)}}}

\def\bra#1{\langle#1\!\mid}
\def\ket#1{\mid\!#1\rangle}
\def\inn#1#2{\langle #1\mid #2\rangle}

\begin{document}

\title{Spin-Orbital Driven Ferroelectricity}

\author{Shan Zhu and You-Quan Li}
\affiliation{
Zhejiang Institute of Modern Physics and
Department of Physics, Zhejiang University, Hangzhou 310027, P. R. China
}

\begin{abstract}
Proposing a general formulism in terms of local coordinates representing
the tilting of the ligands' octahedra,
we evaluate the electric polarization in a chain of transition metal ions
with unpolar octahedron rotation.
We find the orbital ordering produced by the ligands's rotation and the spin order, together,
determines the polarization features,
manifesting
that nonvannishing polarization appears in collinear spin order 
and the direction of polarization is no more restricted in the plane of spin rotation
in cycloidal ordering.
\end{abstract}


\received{\today}

\maketitle

Multiferroics refers to materials in which the spontaneous ferroelectricity and  magnetism coexist,
which makes it possible to control magnetization through electric field or vice versa~\cite{0022-3727-38-8-R01,Eerenstein2006,Cheong2007,Fiebig2009,Wang2009,Physics.2.20}.
Since the discovery of spin-driven multiferroics
$\mathrm{TbMnO_3}$~\cite{Kimura2003} and $\mathrm{TbMn_2O_5}$~\cite{Hur2004},
it has absorbed more and more attentions.
In spiral magnetic insulator, a spin-current mechanism~\cite{PhysRevLett.95.057205}
which is also referred as ``inverse Dzyaloshinskii-Moriya (DM)
interaction''~\cite{PhysRevB.73.094434,PhysRevLett.96.067601}
predicts that the electric polarization is
closely related to the vector spin chirality.
This mechanism fulfills in spiral multiferroics for systems  with high symmetry.
Moreover in collinear spin order system the ferroelectricity is induced by inverse Goodenough-Kanamori mechanism where the exchange
interaction is responsible for the presence of ferroelectricity in Ising chain magnet~\cite{PhysRevLett.100.047601} and $E$-type spin order~\cite{Sergienko2006,Picozzi2007,PhysRevB.78.014403}.
However in $E$-type materials,
the magnitude of electric polarization measured in experiments~\cite{PhysRevB.76.104405,PhysRevB.85.184406}
is about two order of magnitude smaller
than some other theoretical predictions~\cite{Sergienko2006,Picozzi2007,PhysRevB.78.014403}.

We notice in orthorhombic(o)
$o$-$R\mathrm{MnO_3}$ ($R=\mathrm{Ho}$, $\mathrm{Er,\;Tm,\;Yb}$,
and $\mathrm{Lu}$)
with $E$-type spin order
that the tilting of the ligands' octahedra of transition metal ions always takes place.
In recently found ferroaxials~\cite{Johnson2011,Johnson2012,PhysRevLett.108.237201},
both the spin order and special crystal structure
are responsible for the inversion symmetry broken.
In $\mathrm{CaMn_{7}O_{12}}$ ``propeller-like" structure for which the tilted ligands' octahedra take shape,
the axial vector due to the relative rotation between nearby octahedra (we simply call octahedron rotation hereafter)
is essential in the inversion symmetry breaking~\cite{Physics.5.16}.
There has been a phenomenological explanation~\cite{Johnson2012,Physics.5.16} for
the relationships between the electric polarization and the axial vector as well as spin orders,
but there is not yet a microscopic mechanism
to clarify the role that such an octahedron rotation plays in conventional spin-driven multiferroics.
It is therefore obligatory, from a microscopic point of view,
to elucidate the effect of octahedron rotation on spin induced ferroelectricity.

In this letter we investigate how does an occurrence of the unpolar octahedron rotation
affect the final electric polarization in a class of multiferroics.
Starting from microscopic model, we derive an explicit dependence of
the electric polarization on the spin order and the configurations of octahedron rotation,
then discuss the special features by comparing with the case without
the rotation.

\begin{figure}[t]\begin{center}
\includegraphics[scale=0.38]{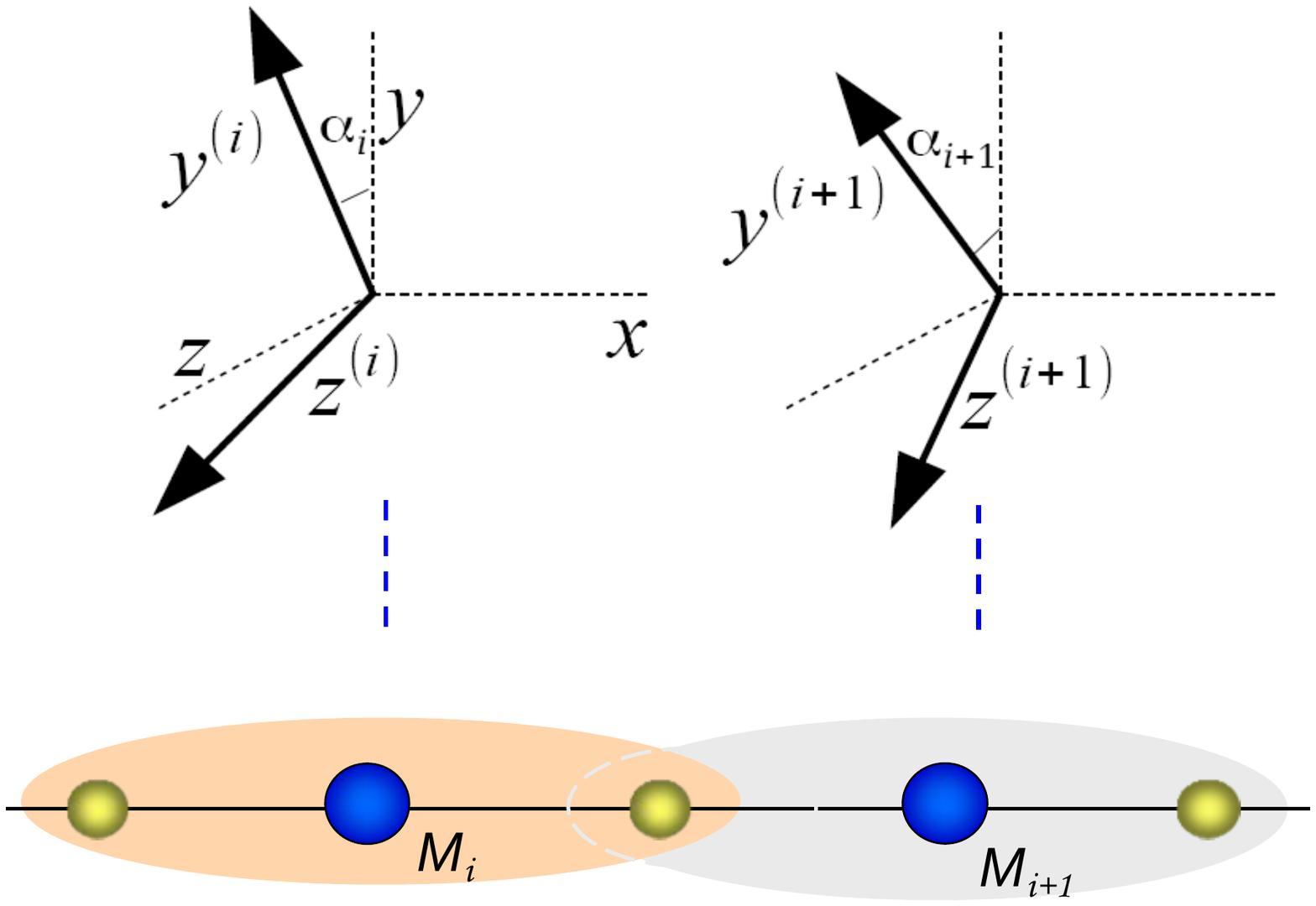}
\vspace{-1mm}
\caption{(Color online)
Schematic illustration of local coordinate frames for the model.
The neighborhood $\mathscr{U}^{(i)}$ shadowed in same color
covers a transition metal ion $M_i$ (large ball)
and its neighboring oxygen atoms (small balls, \ie, a left one and a right one,
they are denoted, respectively, by $\tilde{p}$ and $p$ in the Hamiltonian of this paper
for notation convenience).
One can see, the right oxygen atom in $\mathscr{U}^{(i)}$
is simultaneously the left one in the next neighborhood $\mathscr{U}^{(i+1)}$.
This overlapping in description naturally defines a metric.
The rotation angle $\alpha$ specifies the relation between the global coordinate frame $(x, y, z)$
and a local frame (specified with a superscript $i$ and $i+1$)
that is the same for both $d$ and $p$ states in the same neighborhood.
One needs a nontrivial bundle description in mathematical modeling
in the presence of octahedron rotation in the material.
}
\label{fig:frame}
\end{center}
\end{figure}

To constitute a much more general model, we consider a system consisting of transition metal ions placed in a one dimensional lattice space $L_M$.
Their bonds are bridged through an oxygen atom in between the ions.
Let $L_O$ denote the lattice space where oxygen atoms are placed.
The present model is actually defined on a bundle~\cite{Schutz},
a mathematical concept
that we describe in the following.
In a bipartite lattice $L=L_M \bigcup L_O$,
each lattice point of $L_M$ is attached with a fibre $\mathscr{F}_d$
and that of $L_O$  is attached with a fibre $\mathscr{F}_p$
where $\mathscr{F}_d$ and $\mathscr{F}_p$ stand for the Hilbert space
of $d$ electrons and $p$ electrons respectively.
The structure group $\mathscr{G}$ that connects the nearby fibres is the rotation group.
Then we are in the position to set up a much more generalized Hamiltonian
\begin{eqnarray}
H = \sum_{i,\sigma}H_{\sigma}^{_{(i)}}
    -U\sum_i \boldsymbol{e}_i\cdot\boldsymbol{S}_i,
\label{Hamiltonian}
\end{eqnarray}
with
\begin{eqnarray*}
\Hlocal{i} = V
  \bigl(\pdagger{y } \opd{x y } + \pdagger{z } \opd{x z }
   - \tpdagger{y } \opd{x y } -\tpdagger{z } \opd{x z }\bigr)
         \nonumber\\
   \hspace{1mm} + V_{0}\bigl(\pdagger{x }-\tpdagger{x } \bigr)
  \bigl(\opd{3z ^2-r^2}-\sqrt{3}\opd{x ^2-y ^2}\bigr)+
   \mathrm{H.c.},
 \label{local-terms}
\end{eqnarray*}
where $\sigma=\uparrow,\downarrow$, $i\in L_M$,
$V=t_{pd\pi}$ and $V_{0}=-\frac{1}{2}t_{pd\sigma}$.
The summation in Eq.~(\ref{Hamiltonian})
is taken over the lattice sites that transitional metal ions are placed.
According to octahedral symmetry, the $d$ orbitals of every metal ions are expressed
in the local octahedral basis.
Thus we introduce local coordinates for every transition metal ion.
Precisely, the above $H_{\sigma}^{^{(i)}}$ is defined on the $i$-th neighborhood
$\mathscr{U}^{(i)}$,
in which
$\opd{\nu}$ ($\nu=xy, xz, yz, \mathrm{etc.}$) annihilates
a $d$ electron at site $M_i$ in Fig.~\ref{fig:frame},
$\tpdagger{k}$ ($k=x, y, z $) creates a $p$ electron in the left side of $M_i$, whereas
$\pdagger{k}$  creates one in the right side.
Here we use the superscript ``$(i)$" to emphasis that the corresponding wavefunctions
are expressed in terms of local coordinates in the neighborhood
$\mathscr{U}^{(i)}$.
We emphasis that the quantization axis of spin is taken to be the $z$-axis of local frame rather than the global frame.
The coordinate frame for wavefunction is locally defined on each neighborhood $\mathscr{U}^{(i)}$,
which changes from neighborhood to neighborhood.

In order to visualize the relation between electric polarization and spin order,
we need a final expression in terms of a global coordinate frame,
saying $\boldsymbol{r}=(x, y, z)$.
Hereafter, polarization means electric polarization for simplicity.
Suppose the local coordinate in the neighborhood $\mathscr{U}^{(i)}$
is related to the global coordinate through  a rotation
$\bm{r}^{_{(i)}T} = R(\bm{\alpha}_i)\,\bm{r}^T$
(the superscript ``$T$'' means transpose of matrix, and $R(\bm{\alpha})$ is the fundamental representation of rotation group),
the spin states expressed in local frame and global frame are
connected through the spinor representation of rotation group, which we denote by $R_s(\bm{\alpha})$.
As illustrated in Fig.~\ref{fig:frame},
the $p$-states in the oxygen site between $M_i$ and $M_{i+1}$
can be expressed in terms of
either the local frame of the neighborhood $\mathscr{U}^{(i)}$
or that of the neighborhood $\mathscr{U}^{(i+1)}$.
These two expressions give us a metric
\begin{eqnarray}
\inn{\ip{k,\sigma} }{ \tipp{k'\!,\sigma'}}
    = G^{}_{k\sigma,k'\!\sigma'}.
\label{metric}
\end{eqnarray}
which will appear in the formula of 2nd order perturbation theory.
This metric matrix is  given by
$\bigl(R(\bm{\alpha}_i)\otimes R_s(\bm{\alpha}_i)\bigr)^\dagger
\bigl(R(\bm{\alpha}_{i+1})\otimes R_s(\bm{\alpha}_{i+1})\bigr)$ generally.
We are interested in the case that
the local coordinate in the neighborhood $\mathscr{U}^{(i)}$
is related to the global coordinate through  a rotation of $\alpha_i$
around the $x$-axis, then we have
$R(\alpha)=\mathrm{e}^{\alpha\hat\ell_x}$
where the generator $\hat\ell_x$ is a $3$ by $3$ matrix whose
nonvanishing elements are $(\hat\ell_x)_{23}=-(\hat\ell_x)_{32}=1$,
and $R_s(\alpha)=\mathrm{e}_{}^{i\frac{\alpha}{2}\hat\sigma_x}$
with $\hat\sigma_x$ denoting the Pauli matrix.
For simplicity, the direction $\boldsymbol{e}_{i,i+1}$
that represents the chain direction of transition metal ions
is chosen as $x$-axis.
The metric  $G^{}_{k\sigma,k'\!\sigma'}(\alpha_{i,i+1})$ is characterized by
the relative rotation angle
$\alpha_{i,i+1}=\alpha_{i+1} -\alpha_i$
between the nearest neighbors.
For the octahedral symmetry in $\mathscr{F}_d$,
the rotation with a special angle like $\pi/2$ or its multiplier
is equivalent to the case without rotation.
So the twisting that causes nontrivial rotation is parameterized by $-\pi/4<\alpha_{i,i+1}<\pi/4$.

To obtain the electric polarization
we need to know the orbital components of the ground state.
The second part of Eq.~(\ref{Hamiltonian}) describes the
$d$ electron of spin $\boldsymbol{S'}_i$ coupling with the local magnetic moment
along the direction
$\boldsymbol{e}^\prime_i=(\sin\theta'_i\cos\phi'_i, \sin\theta'_i\sin\phi'_i, \cos\theta'_i )$.
Here we use prime to emphasis that the coordinate parameters are defined with
respect to the local coordinate frame.
Like in Ref.~\cite{PhysRevLett.95.057205}, we consider the case that the
strength of spin-orbit coupling is the largest energy scale of the system.
Then the first part of Eq.~(\ref{Hamiltonian}) can be applied as
a perturbation in the Hilbert subspace for the ground states of the second one.
Diagonalizing the $-U\boldsymbol{e}^\prime_i\cdot\boldsymbol{S'}_i$ term,
one obtain
\begin{eqnarray}
\ket{P_i} &  = \sin\frac{\theta_i'}{2}\ket{a^{(i)}}
  + \mathrm{e}^{i\phi_i'} \cos\frac{\theta_i'}{2}\ket{b^{(i)}},
   \nonumber\\[1mm]
\ket{\tilde{P}_i} &= \cos\frac{\theta_{i}'}{2}\ket{a^{(i)}}
  -\mathrm{e}^{i\phi_i'}\sin\frac{\theta_i'}{2}\ket{b^{(i)}},
\label{parallel}
\end{eqnarray}
where
$\ket{a^{(i)}}=(\ket{\id{x y}{\uparrow}} + \ket{\id{y z}{\downarrow}} + i\ket{\id{z x}{\downarrow}}  )/\sqrt{3}$
and
$\ket{b^{(i)}}=(\ket{\id{x y}{\downarrow}} - \ket{\id{y z}{\uparrow}} + i\ket{\id{z x}{\uparrow}}  )/\sqrt{3}$
are the $\Gamma_7$ doublet defined in the local coordinate
of the neighborhood $\mathscr{U}^{(i)}$.
Since
$\bra{P_i}\boldsymbol{\sigma}\ket{P_i}=\boldsymbol{e}'_i$
and
$\bra{\tilde{P}_i}\boldsymbol{\sigma}\ket{\tilde{P}_i}=-\boldsymbol{e}'_i$,
$\ket{P_i}$ and $\ket{\tilde{P}_i}$ are called
parallel and anti-parallel states respectively.
The eigenenergy of $\ket{\tilde{P}_i}$ and $\ket{P_i}$
are $U/3$ and $-U/3$ respectively.
For convenience in calculation, one can abbreviate the aforementioned parallel state as
$\ket{P_i}=\sum_{\nu\sigma}\iB{\nu}{\sigma}\ket{\id{\nu}{\sigma}}$
with the corresponding coefficients:
$\iB{y z }{\downarrow}
=\iB{z x }{\downarrow}
=\iB{x y }{\uparrow}
=\sin\frac{\theta_i'}{2}/\sqrt{3}$,
and
$-\iB{y z }{\uparrow}
=-i\iB{z x }{\uparrow}
=\iB{x y }{\downarrow}
=\mathrm{e}^{i\phi_i'}\cos\frac{\theta_i'}{2}/\sqrt{3}$.

\begin{figure}[h!]
\begin{center}
\includegraphics[scale=0.40]{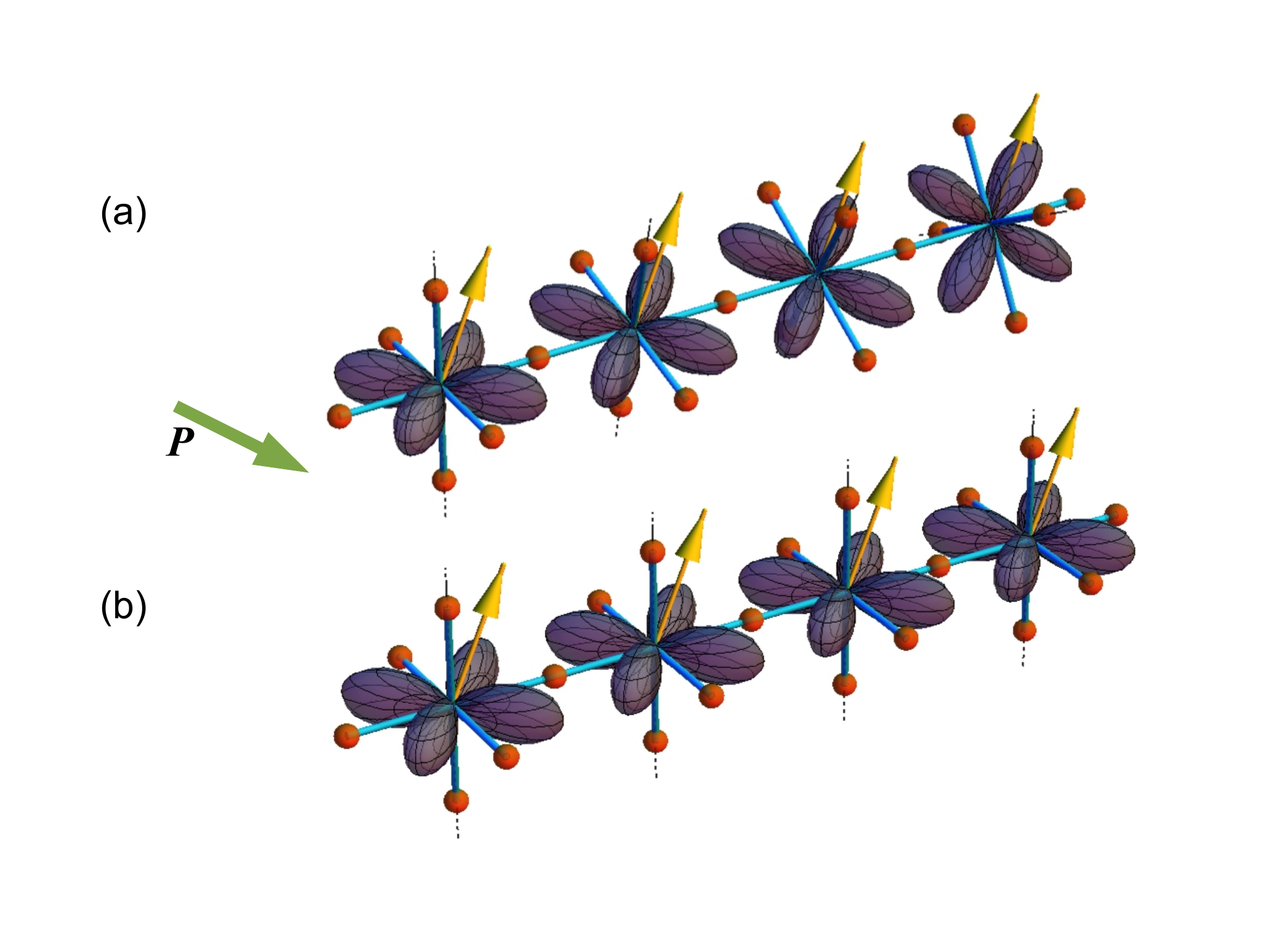}
\caption{
(Color online)
Depiction of the spin order (yellow arrows), orbital ordering ($d_{xz}$) in the presence of octahedron rotation along $x$-axis  and
the resulting total polarization (green arrow).
For collinear spin order,
the total polarization $\boldsymbol{P}$ turns to appear in the present of octahedron rotation (a)
though it is known~\cite{PhysRevLett.95.057205} to vanish in the absence of octahedron rotation (b).
Here the non-uniform orbital ordering produced by octahedron rotation (a) plays an important role.}
\label{fig:spin-orbital}
\end{center}
\end{figure}

In the second order perturbation for the degenerate ground states $\ket{P_i}$
and $\ket{P_{i+1}}$, only the terms
$\Hlocal{i}$ and $\Hlocal{i+1}$ are of relevance
because they describe hopping to or from the oxygen atom in between $M_i$ and $M_{i+1}$.
From the secular equation
\begin{eqnarray}
\sum_{j'}(K_{j,j'} - {\mathcal E}\delta_{j,j'})C_{j'}=0, \quad (j,j'= i, i+1),
 \label{secular}
\end{eqnarray}
for the $2$ by $2$ matrix
\begin{align*}
K_{j,j'}=\sum_{k,\sigma,k'\sigma'}
  \frac{
   \bra{P_j} \Hlocal{i}\ket{\ip{k\sigma}} G_{k\sigma,k'\sigma'}\bra{\tipp{k'\sigma'}} \Hlocal{j'} \ket{ P_{j'} }
  }
  {-\Delta },
\end{align*}
where $\Delta$ is the energy difference between the $p$ and $d$ orbitals,
we obtain  two roots (eigenenergies)
${\mathcal E}^{_I}=\frac{2}{3}V^2(1-\left|b\right|)$ and ${\mathcal E}^{_{I\!I}}=\frac{2}{3}V^2(1+\left|b\right|)$
and  the corresponding eigenvectors
$( C^{_I}_i , C^{_I}_{i+1}) = \frac{1}{\sqrt{2}}(-\frac{b}{\left|b\right|},1)$;
$( C^{_{I\!I}}_i , C^{_{I\!I}}_{i+1}) = \frac{1}{\sqrt{2}}(\frac{b}{\left|b\right|},1)$
where
$b=\mu_{i+1}^{\dagger}\mathrm{e}^{-i\alpha_{i,i+1}\hat\sigma_{x}}\mu_i$,
with
$\mu_{i}^\dagger = (\cos(\frac{\theta_i}{2}),
\sin(\frac{\theta_i}{2})e^{i\phi_i})$.
%
Then we have two eigenkets
\begin{align}
\ket{\Psi_i^a} & =  C^{a}_{i} \ket{P_i}+ C^{a}_{i+1} \ket{P_{i+1}}
 \nonumber\\
  & + \frac{V}{\Delta}\sum_{k,\sigma}\bigl(
   -D^{a}_{k\sigma,i}\ket{\ip{k\sigma}}+ \tilde{D}^{a}_{k\sigma,i+1}\ket{\tipp{k\sigma}}
   \bigr),
 \label{eigenket}
\end{align}
where $a=\emph{\small I}, \emph{\small I\!I}$ label the two lowest energy states.
Those coefficients in Eq.~(\ref{eigenket}) are
$D^{a}_{x \sigma,i}= \tilde D^{a}_{x \sigma,i+1}=0$,
$D^{a}_{y \sigma,i}=C^{a}_{i}\iB{x y }{\sigma}$,
$D^{a}_{z \sigma,i}=C^{a}_{i}\iB{x z }{\sigma}$,
$\tilde D^{a}_{y \sigma,i}=C^{a}_{i}\iBp{x y }{\sigma}$,
$\tilde D^{a}_{z \sigma,i}=C^{a}_{i}\iBp{x z }{\sigma}$.
Note that the
$\ket{P_i}$, $\ket{\ip{k\sigma}}$,
$\ket{P_{i+1}}$ and $\ket{\tipp{k\sigma}}$ in Eq.~(\ref{eigenket})
are expressed in terms of local coordinates.
In the presence of octahedron rotation,
the orbital state implied in $\ket{P_i}$ differs in different site $i$,
which is illustrated in Fig.~\ref{fig:spin-orbital}.
Thus the orbitals at each site of transition metal are no more uniformly ordered along the chain.
Such an orbital ordering is characterized by the twisting angle $\alpha_{i,i+1}$ along $x$-axis.
Only for  either parallel or antiparallel spin orders meanwhile $\alpha_{i,i+1}=0$,
the state (\ref{eigenket}) becomes an eigenstate of the operator of inversion transformation.
If $\alpha_{i,i+1}\neq 0$, however,
the inversion symmetry is still broken even for collinear spin orders,
and hence nonvanishing polarization is expected  to appear then.

Now we are in the position to evaluate the induced polarization.
The overlapping integrals of $d$ and $p$ orbitals contributed
in the calculation of the expectation value for $e\boldsymbol{r}$ are~\cite{PhysRevLett.95.057205},
\begin{align}
\int\id{y z }{\sigma} y^{(i)} \tipp{k,\sigma'}\mathrm{d}^3\boldsymbol{r}
 &=
G^{}_{z \sigma,k\sigma'}(\alpha_{i,i+1})A,
 \label{ImetricP}
\end{align}
where $A=\int\opd{y z }(\boldsymbol{r})y ^{(i)} p_{z,\sigma}^{(i)}(\boldsymbol{r})\mathrm{d}^3\boldsymbol{r}.$
The same is true for $y\rightarrow z, z\rightarrow y$ in the above.
Let us discuss the polarization for the cases of  one hole as well as two holes in the one-electron picture.
In the one hole case, we have the hole in the lowest energy state,
$\ket{\Psi_i^{_I}}$.
The polarization is given by
$\boldsymbol{P}_i = e\bra{\Psi_i^{_I}} \boldsymbol{r} \ket{\Psi_i^{_I}}/\inn{\Psi_i^{_I}}{\Psi_i^{_I}}$.
If there are two holes that will occupy the two lowest states respectively, the polarization is defined as
$\boldsymbol{P}_i = e\bra{\Psi_i^{_I}} \boldsymbol{r} \ket{\Psi_i^{_I}}/\inn{\Psi_i^{_I}}{\Psi_i^{_I}}
  + e\bra{\Psi_i^{_{I\!I}}} \boldsymbol{r} \ket{\Psi_i^{_{I\!I}}}/\inn{\Psi_i^{_{I\!I}}}{\Psi_i^{_{I\!I}}}$.
Consequently, we have
\begin{align}
\boldsymbol{P}_{i} =\lambda\, \boldsymbol{e}_{i,i+1}\times(\mathrm{e}^{-\alpha_{i,i+1}\hat\ell_x}\boldsymbol{e}_{i}
\times \mathrm{e}^{\alpha_{i,i+1}\hat\ell_x}\boldsymbol{e}_{i+1}),
\label{twoHP}
\end{align}
where
$\lambda=\lambda_0 \cos(\alpha_{i, i+1})$.
Here
$\lambda_0 = \frac{1}{3}eAV/(\Delta |b| )$ and
$\lambda_0 = -\frac{4}{9}eA(V/\Delta)^{3} $ for one hole and two holes, respectively.
%
%
Note that
$\mathrm{e}^{-\alpha_{i,i+1}\hat\ell_x}\boldsymbol{e}_{i}
\times \mathrm{e}^{\alpha_{i,i+1}\hat\ell_x}\boldsymbol{e}_{i+1}$ in Eq.~(\ref{twoHP})
is happened to be the spin current for a tilt Heisenberg superexchange model~\cite{PhysRevB.84.205123}.

We can perceive several features from Eq.~(\ref{twoHP}).
It is no more the spin chirality itself~\cite{PhysRevLett.95.057205,Jia2007}
but the spin order together with the orbital ordering that determine the direction of polarization.
Additionally, the twisting angle modulates
the magnitude of polarization via $\cos\left(\alpha_{i,i+1}\right)$,
thus the order of magnitude of ferroelectricity keeps unchanged compared with Ref.~{\cite{PhysRevLett.95.057205}}.
Clearly,
the above formula (\ref{twoHP}) reduces to the result of Ref.~{\cite{PhysRevLett.95.057205}}
for an uniform orbital ordering (\ie, in the absence of octahedron rotation).
In the following we discuss the application of Eq.~(\ref{twoHP})
to various spin orders.

\begin{figure}
\begin{center}
\includegraphics[scale=0.38]{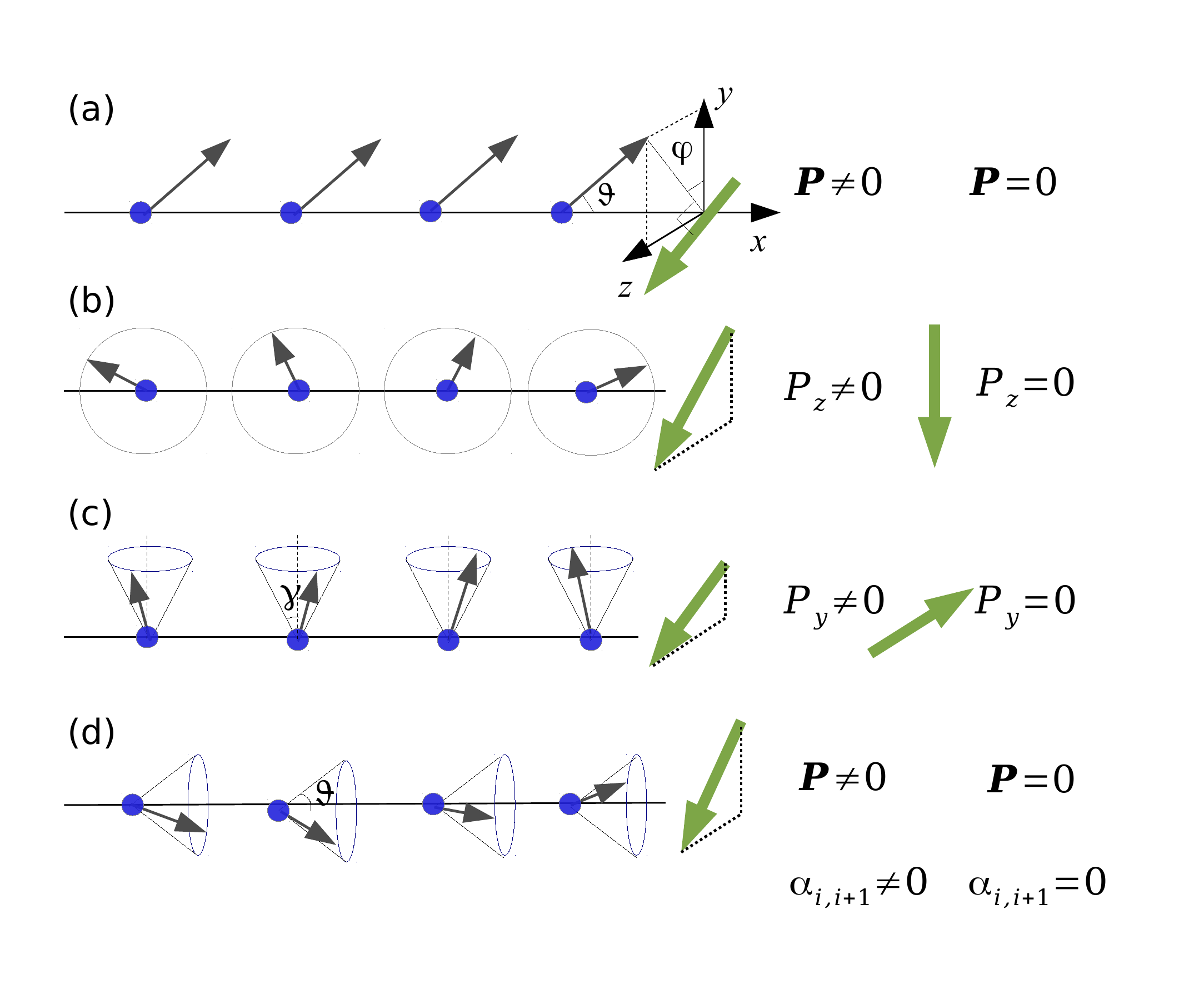}
\caption{
(Color online)
Illustration of the relations between total polarization $\boldsymbol{P}$ and various magnetic orders,
such as collinear (a), cycloidal (b), transverse conical (c), and
longitudinal conical (d) orders.
Here the black arrows denote the spin direction at every site
and the green one represents the induced polarization.
As a comparison, the induced polarization for the special case
without twisting $\alpha_{i,i+1}=0$
are also plotted at the right accordingly.
It is worthwhile to emphasize that,
for (a) and (d),
the polarization vanished in the uniform case $\alpha_{i,i+1}=0$
turns to appear in the nonuniform case $\alpha_{i,i+1}\neq 0$.}
\label{fig:compare}
\end{center}
\end{figure}

\paragraph{Nonvanishing polarization appearing in collinear orders:}
Let us consider a ferromagnetic order, \ie, parallel spin aligning along
the direction
$\left(\mathrm{\cos\vartheta,\:\sin\vartheta\cos\varphi,\:\sin\vartheta\sin\varphi}\right)$
where $\vartheta$ denotes the zenith angle from the positive $x$-axis and
$\varphi$ denotes the azimuthal angle in the $y$-$z$ plane from the $y$-axis as shown in  Fig.~\ref{fig:compare}(a).
Equation~(\ref{twoHP}) gives rise to
\begin{align}
\boldsymbol{P}_{i}=\lambda_0\sin2(\alpha_{i,i+1})\sin2\vartheta\,\bigl(0,\; -\sin\varphi,\; \cos\varphi\bigr),
\end{align}
that implies the appearance of nonvanishing polarization $\boldsymbol{P}_i$ for the parallel spin order. For the case of antiparallel neighboring spins, the polarization $\boldsymbol{P}_i$ in Eq.~(\ref{twoHP}) will have an opposite sign.
Actually, as we can see from the above equation
that the $\boldsymbol{P}_i$ is perpendicular to the spin direction as well as $\boldsymbol{e}_{i,i+1}$ which is also demonstrated in Fig.~\ref{fig:compare}(a).
Our formula also manifests that the induced polarization $\boldsymbol{P}_i$ in collinear case is in the same order of magnitude with that for the spiral insulators~\cite{PhysRevLett.95.057205}.
Thus the combination of collinear spin order and non-uniform orbital ordering due to octahedron rotation (see Fig.~\ref{fig:spin-orbital}) will break the inversion symmetry.
We notice that in $o$-$R\mathrm{MnO_3}$
with $E$-type spin order the $\boldsymbol{P}_i$ is also perpendicular to the spin direction~\cite{Sergienko2006,Picozzi2007,PhysRevB.78.014403,PhysRevB.76.104405,PhysRevB.85.184406}.

\paragraph{Out-of-plane tilting of the polarization existing in cycloidal magnetic order:}
We consider the cycloidal magnetic order Fig.~\ref{fig:compare}(b),
namely, $\boldsymbol{e}_{i}=(\cos\vartheta_{i},\:\sin\vartheta_{i},\:0)$.
Here the spin lies in the $x$-$y$ plane and $\vartheta_{i}$ denotes the angle between
$x$-axis and the spin on the $i$th site.
Equation~(\ref{twoHP}) becomes
\begin{align}
\left\{\begin{array}{l}
      (\boldsymbol{P}_{i})_x =   0, \\[2mm]
      (\boldsymbol{P}_{i})_y =  \lambda_0\cos^2(\alpha_{i,i+1})\sin(\phi_{i}-\phi_{i+1}), \\[2mm]
      (\boldsymbol{P}_{i})_z =  \frac{1}{2}\lambda_0\sin 2(\alpha_{i,i+1}) \sin (\vartheta_{i+1}+\vartheta_{i}).
\end{array}\right.
\label{polarizationspiral}
\end{align}
If there are no octahedron rotation, \ie, $\alpha_{i,i+1}=0$, the above polarization formula reduces to the one of Ref.~\cite{PhysRevLett.95.057205},
which means the polarization is just along $y$-axis.
Whereas, in the presence of octahedron rotation $\alpha_{i,i+1} \neq 0$,
Eq.~\eqref{polarizationspiral} tells
us that there will be nonzero $z$-component in the induced polarization,
which is shown in Fig.~\ref{fig:compare}(b).
In $\mathrm{Cu}_{3}\mathrm{Nb}_{2}\mathrm{O}_{8}$~\cite{Johnson2011} the polarization is nearly $14^\circ$
to the $z$-axis,
which can be interpreted by our result (\ref{polarizationspiral})
that implies the polarization is not restricted in the plane of spin rotation.
There are two kinds of local crystal field on $\mathrm{Cu}$ from neighboring $\mathrm{O}$
that are $\mathrm{CuO_4}$ square-plane and square-pyramid.
These two local crystal fields play an important role in the presence of out-of-plane polarization.

\paragraph{Variations in conical magnetic orders:}
Let us discuss the transverse and longitudinal
conical order respectively.
(i) For transverse conical order that is expressed as
$\boldsymbol{e}_{i}^T=(\sin\gamma\cos\psi_{i},\,\cos\gamma,\, \sin\gamma\sin\psi_{i})$
where $\psi_{i}$ denotes the azimuthal angle and $\gamma$ the polar
angle with respect to the generatrix paralleling to the $y$-axis,
the polarization we obtained is given by
$
\boldsymbol{P}_{i}^T\propto \sin^{2}\gamma\sin(\psi_{i,i+1})(0,\,\frac{1}{2}\sin2(\alpha_{i,i+1}),\,
-\cos^{2}(\alpha_{i,i+1}))+\sin2\gamma(\cos\psi_{i+1}-\cos\psi_i)(0,\,
\frac{1}{2}\cos^{2}(\alpha_{i,i+1}),\,\frac{1}{4}\sin2(\alpha_{i,i+1}))
$
where $\psi_{i,i+1}=\psi_{i+1}-\psi_{i}$.
Here we need to figure out that the total polarization $\boldsymbol{P}$ has
$y$-component, which is shown in Fig.~\ref{fig:compare}(c).
(ii) For the longitudinal conical order which is expressed as
$
\boldsymbol{e}_{i}^{L}=(\cos\vartheta,\:\sin\vartheta\sin\varphi_{i},\:\sin\vartheta\cos\varphi_{i})
$
where $\varphi_{i}$ denotes the azimuthal  angle and $\vartheta$ the polar
angle with respect to $x$-axis as the conical generatrix.
The polarization for the longitudinal conical order is given by
$\boldsymbol{P}_{i}^L\propto\frac{1}{2}\sin2\vartheta\cos\left(\alpha_{i,i+1}
\right)(0,\,\sin(\alpha_{i,i+1}+\varphi_{i})-\sin(-\alpha_{i,i+1}
 +\varphi_{i+1}),\,\cos(\alpha_{i,i+1}+\varphi_{i})-\cos(-\alpha_{i,i+1}+
  \varphi_{i+1}))$.
The  total polarization $\boldsymbol{P}$ is not zero
unless one consider the uniform case~\cite{PhysRevLett.95.057205}
of $\alpha_{i,i+1}=0$,
which is shown in Fig.~\ref{fig:compare}(d).
Clearly, when $\vartheta=\pi/2$ the longitudinal conical order turns to be
the screw order,
$\boldsymbol{e}_{i}=(0,\,\cos\varphi_{i},\:\sin\varphi_{i})$
and the polarization vanishes then.
We predict that the polarization will be in the $y$-$z$ plane
for both longitudinal and transverse conical spin orders
as long as there is an octahedron rotation along magnetic propagation $x$-axis.

In summary, we proposed a microscopic formulism in terms of local coordinates to calculate the electric polarization for a class of multiferroics in which the neighboring ligands' octahedra twist.
We obtained an explicit expression
in which both the spin order and the orbital ordering produced by the octahedron rotation
determine the appearance of polarization.
This is due to the spin order and the orbital ordering, together,
break the inversion symmetry.
We find that the nonvannishing polarization appears in collinear spin order
and it is in the same order of magnitude as in spiral insulator.
In some special lattice structure,
we find that the electric polarization in collinear case can be in the same order of magnitude with the spiral multiferroics, and it is no more restricted in the plane of spin rotation in cycloidal case.

We thank R.B. Tao and F.C. Zhang for helpful discussions.
The work is supported by NSFCs (grant No.11074216 \& No.11274272)
and the Fundamental Research Funds for the Central Universities of China.

\bibliography{polarization1}

\end{document}